\documentclass[%
 reprint,
 amsmath,amssymb,
 aps,prb
]{revtex4-1}

\usepackage{graphicx}
\usepackage{dcolumn}
\usepackage{bm}
\usepackage{hyperref}
\usepackage{float,epstopdf}

\begin{document}

\preprint{APS/123-QED}

\title{Generalized analytical solutions and experimental confirmation of complete synchronization in a class of mutually-coupled simple nonlinear electronic circuits}
\author{G. Sivaganesh}
\affiliation{%
Department of Physics, Alagappa Chettiar College of Engineering $\&$ Technology, Karaikudi, Tamilnadu-630 004, India\\
 }%
\author{A. Arulgnanam}
 \email[Corresponding author: ]{gospelin@gmail.com}
\affiliation{%
Department of Physics, St. John's College, Palayamkottai, Tamilnadu-627 002, India\\
 }%
\author{A. N. Seethalakshmi}
\affiliation{%
Department of Physics, The M.D.T Hindu College, Tirunelveli, Tamilnadu-627 010, India\\
 }%

\date{\today}

\begin{abstract}
In this paper, we present a novel explicit analytical solution for the normalized state equations of mutually-coupled simple chaotic systems. A generalized analytical solution is obtained for a class of simple nonlinear electronic circuits with two different nonlinear elements. The synchronization dynamics of the circuit systems were studied using the analytical solutions. the analytical results thus obtained have been validated through numerical simulation results. Further, we provide a sufficient condition for synchronization in mutually-coupled, second-order simple chaotic systems through an analysis on the eigenvalues of the difference system. The bifurcation of the eigenvalues of the difference system as functions of the coupling parameter in each of the piecewise-linear regions, revealing the existence of stable synchronized states is presented. The stability of synchronized states are studied using the {\emph{Master Stability Function}}. Finally, the electronic circuit experimental results confirming the phenomenon of complete synchronization in each of the circuit system is presented. 

\begin{description}

\item[PACS numbers]
05.45.Gg; 05.45.Xt;
\end{description}

\end{abstract}

\maketitle


\section{INTRODUCTION}

The phenomenon of chaos synchronization plays a key role in the application of secure transmission of signals \cite{Ogorzalek1993,Lakshmanan1994}. Several chaotic systems have been studied for synchronization following the {\emph{Master-Slave}} method introduced by Pecora and Carroll \cite{Pecora1990}. A complete study on the different types of synchronization was presented by Boccaletti {\emph{et al }\cite{Boccaletti2002}. The stability of the synchronized states becomes equally important as compared to the type of synchronization observed in the coupled system. The {\emph{Master Stability Function}}(MSF) approach introduced by Pecora and Carroll provided the necessary condition for synchronization of coupled identical chaotic systems \cite{Pecora1998}. The MSF approach has been used to study the stability of the synchronized states in certain unidirectionally-coupled simple chaotic systems \cite{Liang2009,Sivaganesh2016,Sivaganesh2017}. Chaos synchronization has been observed in certain mutually-coupled autonomous chaotic systems \cite{Chua1992,Wagemakers2007}. In this paper, we discuss the phenomenon of complete synchronization in certain diffusively or mutually-coupled identical simple non-autonomous chaotic systems. The chaotic systems chosen for our study belongs to the class of series and parallel LCR circuits with two different piecewise linear nonlinear elements.\\

Explicit analytical solutions to unidirectionally coupled simple chaotic systems and strange non-chaotic systems have been reported recently in literature \cite{Sivaganesh2015,Sivaganesh2016,Sivaganesh2017}. Analytical technique of that kind have been extended to develop solutions for mutually-coupled chaotic systems. The mechanism of chaos synchronization in mutually-coupled simple nonlinear electronic circuit systems exhibiting chaotic attractors have not been studied analytically in the literature. In the present study, we report a novel analytical method for studying the synchronization phenomenon observed in mutually-coupled chaotic systems. The analytical solutions have been developed for two different class of chaotic systems each with two different nonlinear elements. The analytical solutions have been used to obtain the unsynchronized and synchronized states of the coupled systems. Further a simple method to identify the stable of synchronization in mutually-coupled chaotic systems is proposed. The method is based on a study of the eigenvalues of the difference systems corresponding to the mutually-coupled systems. A similar study has been performed in unidirectionally coupled simple chaotic systems to identify the phenomenon of complete synchronization \cite{Sivaganesh2015,Sivaganesh2016a,Sivaganesh2017}. However, in this paper we confirm that the study on the eigenvalues of the difference system of mutually-coupled chaotic systems can provide a sufficient condition for the synchronization of the systems. The reliability of the results thus obtained have been confirmed by comparing it with the {\emph{Master Stability Functions}} obtained for the corresponding systems which provides the necessary condition for complete synchronization. Further, the stability of the fixed points corresponding to the difference system is presented to confirm the convergence of the trajectories within the synchronization manifold. We discuss the complete synchronization phenomenon in four different simple circuit systems namely, the {\emph{Murali-Lakshmanan-Chua}}, the {\emph{Variant of Murali-Lakshmanan-Chua}}, series LCR circuit with a {\emph{simplified nonlinear element}} and the parallel LCR circuit with {\emph{simplified nonlinear element}} circuits. The circuits systems are categorized under two classes as the series LCR circuit with two different types of nonlinear elements and the parallel LCR circuit with two different types of nonlinear elements. \\

This paper is organized as follows. In Sec. \ref{sec:2} we explain the two different nonlinear elements and their role in the series and parallel LCR circuits. The normalized state equations of the circuit systems which are mutually-coupled are also presented. In Sec. \ref{sec:3} we present generalized explicit analytical solutions to the mutually-coupled series and parallel LCR chaotic systems exhibiting complete synchronization in their dynamics. Further, the unsynchronized and synchronized states of the coupled systems are studied using the analytical solutions obtained. Section \ref{sec:4} deals with the numerical study on complete synchronization and its stability using the MSF approach for series LCR circuits with different nonlinear elements while Sec. \ref{sec:5} with that of the parallel LCR circuits. The experimental results are presented for each case presented in Sec. \ref{sec:4} and \ref{sec:5}. Finally, a discussion on the sufficient condition for synchronization in mutually-coupled second-order chaotic systems  based on the eigenvalues of the difference system is presented in Sec. \ref{sec:6}.\\

\section{Circuit Equations}
\label{sec:2}

A sinusoidally forced series LCR circuit with a piecewise-linear nonlinear element $N_R$ connected parallel to the capacitor is as shown in Fig. \ref{fig:1}. The nonlinear element $N_R$ can be of two types as shown in Fig. \ref{fig:2}. The first type of nonlinear element shown in Fig. \ref{fig:2}(a) is the {\emph{Chua's diode}} introduced by Chua {\emph{et al}}, is constructed using two operational amplifiers and six linear resistors. The corresponding $(v-i)$ characteristics shown in Fig. \ref{fig:2}(b) represents two negative outer slope regions and one negative inner slope region. The state equations of the circuit with the {\emph{Chua's diode}} as the nonlinear element called as the {\emph{Murali-Lakshmanan-Chua}} circuit is given as
\begin{figure}
\begin{center}
\includegraphics[scale=0.66]{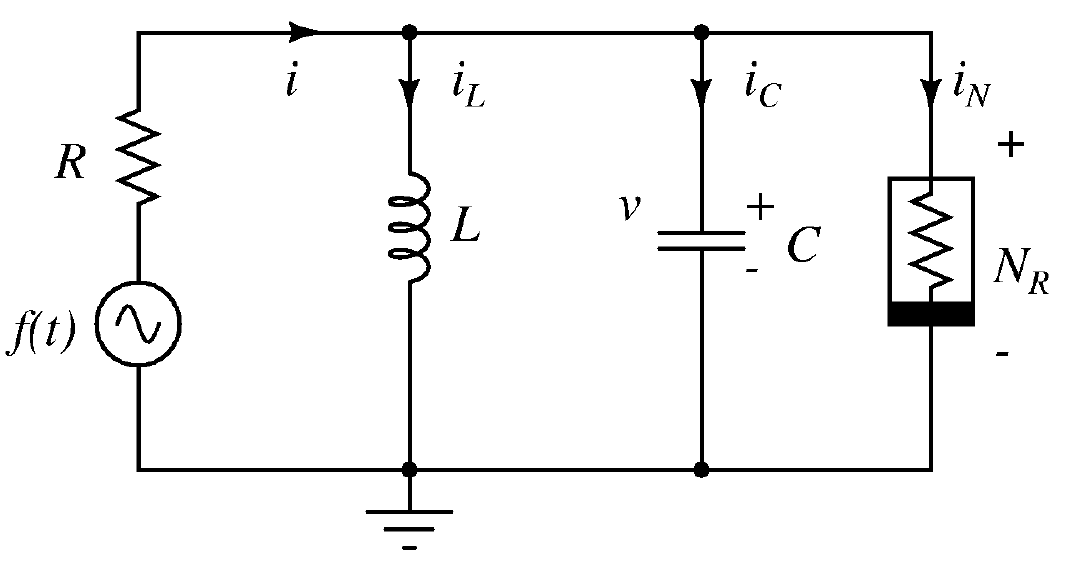}
\caption{Schematic circuit realization of the sinusoidally forced series LCR circuit with the nonlinear element $N_R$ connected parallel to the capacitor.}
\label{fig:1}
\end{center}
\end{figure}
\begin{subequations}
\begin{eqnarray} 
C {dv \over dt } & = & i_L - g(v), \\
L {di_L \over dt } & = & -R i_L - R_s i_L - v + F sin( \Omega t),
\end{eqnarray}
\label{eqn:1}
\end{subequations}
where $g(v)$ is the mathematical form of the voltage-controlled, piecewise-linear resistor given by
\begin{eqnarray}
g(v) =
\begin{cases}
{G_b}v+({G_a}-{G_b}) & \text{if $v \ge 1$}\\
{G_a}v & \text{if $|v|\le 1$}\\
{G_b}v-({G_a}-{G_b}) & \text{if $v \le -1$}
\end{cases}
\label{eqn:2}
\end{eqnarray}
\begin{figure}
\begin{center}
\includegraphics[scale=0.5]{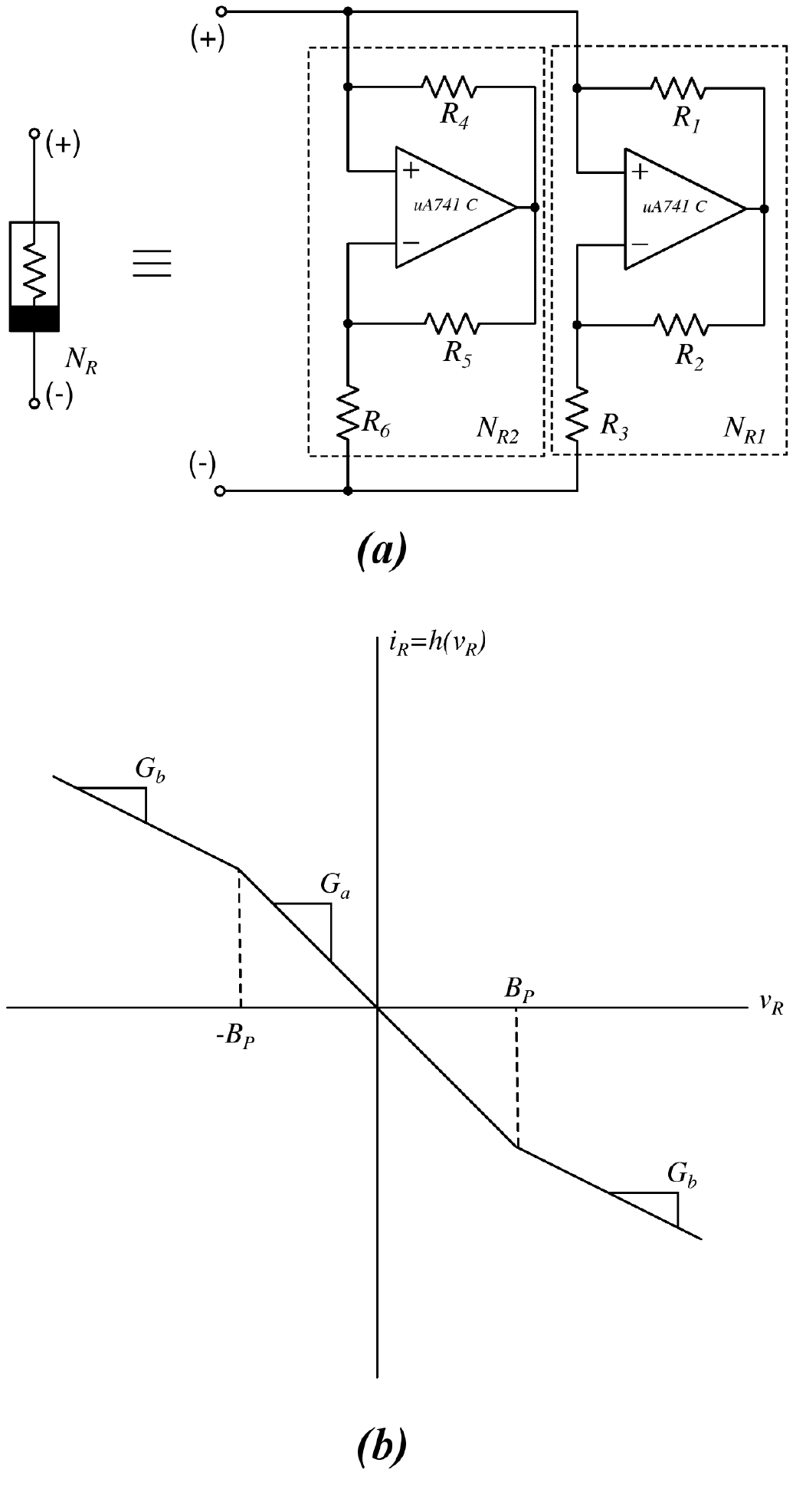}
\caption{(a) Schematic realization of the {\emph{Chua's diode}} constructed from two operational amplifiers and six linear resistors and its (b) $(v-i)$ characteristics with one negative inner slope $G_{a} = -0.76$ mS, two negative outer slopes $G_{b} = -0.41$ mS and the break points $B_{p} = \pm~1.0$ V, respectively.}
\label{fig:2}
\end{center}
\end{figure}
\begin{figure}
\begin{center}
\includegraphics[scale=0.5]{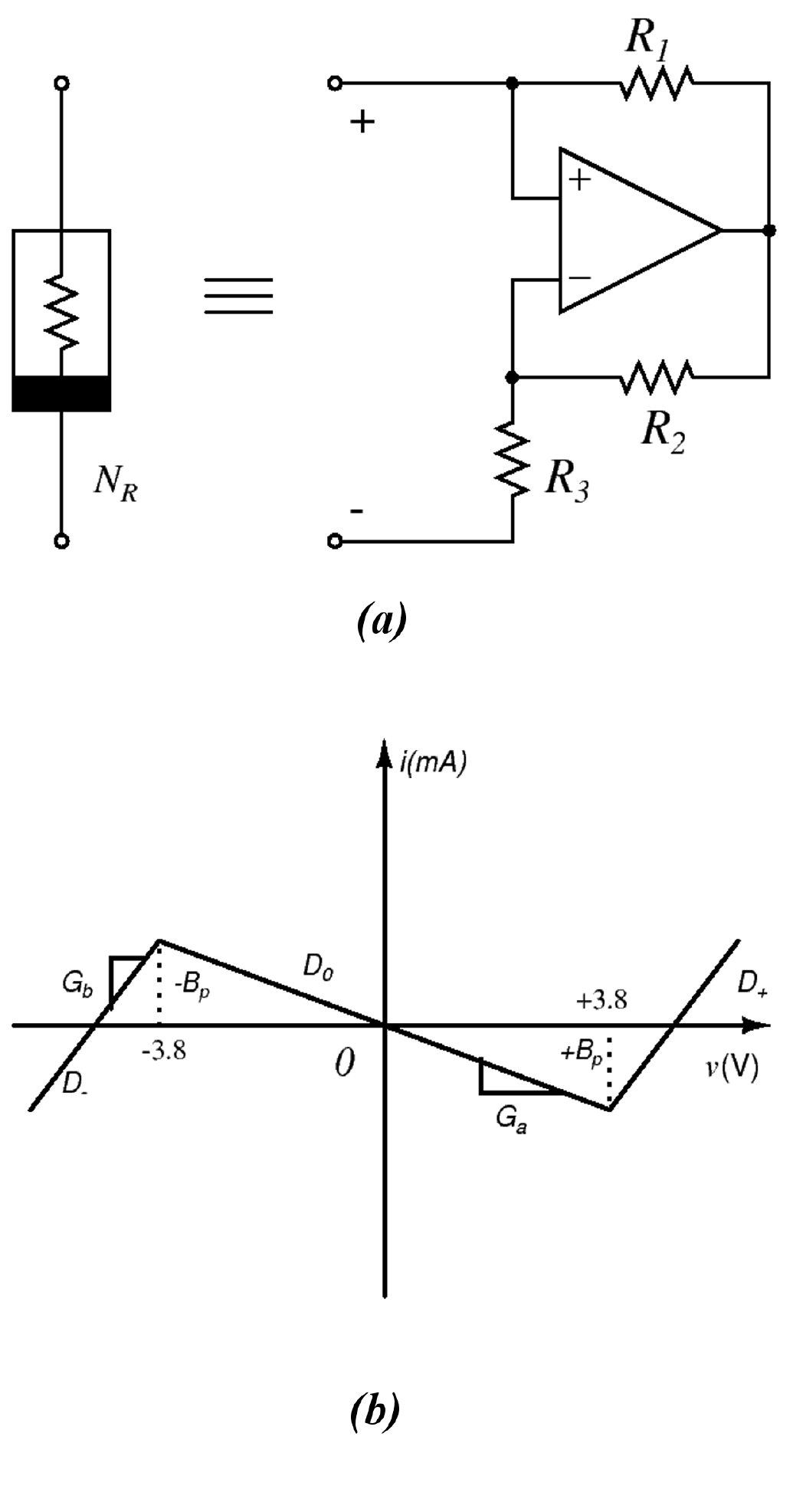}
\caption{(a) Schematic realization of the {\emph{simplified nonlinear element}} constructed using one operational amplifier and three linear resistors and its (b) $(v-i)$ characteristics with a negative inner slope $G_{a} = -0.56$ mS, two positive outer slopes $G_{b} = +2.5$ mS and the break points $B_{p} = \pm~3.8$ V, respectively.}
\label{fig:3}
\end{center}
\end{figure}
The second type of nonlinear element called the {\emph{simplified nonlinear element}} is as shown in Fig. \ref{fig:3}, introduced by Arulgnanam {\emph{et al}}, has been constructed using only one operational amplifier and three linear resistors. This nonlinear element has been identified to be the simplest piecewise-linear element constructed using fewer circuit elements. The state equations of a series LCR circuit with a {\emph{simplified nonlinear element}} connected parallel to the capacitor is given by
\begin{subequations}
\begin{eqnarray} 
C {dv \over dt } & = & i_L - g(v), \\
L {di_L \over dt } & = & -R i_L + F sin( \Omega t),
\end{eqnarray}
\label{eqn:3}
\end{subequations}
The function $g(v)$ is represented the same as given by Eq. \ref{eqn:2} except that the circuit parameters $G_a$ and $G_b$ are different for the two elements. Thus the nonlinear element $N_R$ determines the chaotic dynamics of the circuit for a proper choice of the other circuit parameters. The state equations of the series LCR circuits with different nonlinear elements given by Eqs. \ref{eqn:1} and \ref{eqn:3} can be written in a general form as \\
\begin{subequations}
\begin{eqnarray}
\dot x & = & y - g(x), \\
\dot y & = & - \sigma  y - \beta x + f sin( \theta)  ,\\
\dot \theta & = & \omega,
\end{eqnarray}
\label{eqn:4}
\end{subequations}
The piecewise linear function $g(x)$ representing the three-segmented piecewise nonlinear element is given by,
\begin{equation}
g(x) =
\begin{cases}
bx+(a-b) & \text{if $x\ge 1$}\\
ax & \text{if $|x|\le 1$}\\
bx-(a-b) & \text{if $x\le -1$}
\end{cases}
\label{eqn:5}
\end{equation}
where, $\sigma = (\beta + \nu \beta)$ and $ \beta = (C/LG^2)$, $ \nu = GR_s$, $ a  = G_a/G$,  $ b = G_b/G$, $f_1 = (F_1 \beta/B_p)$, $\omega_1 = (\Omega_1 C/G)$, $ G = 1/R$. The values of the normalized parameters of the system depends upon the circuit parameters. For the circuit with a simplified nonlinear element, $\nu = 0$ and hence $\sigma = \beta$.\\  
The circuit said above can be mutually-coupled with another circuit having the same nonlinear element. In this case, the normalized state variables of the two systems can be represented as $(x,y))$ and $(x^{'}, y^{'})$, respectively. The normalized state equations of the mutually-coupled systems under coupling of the $x$ variables is given as
\begin{subequations}
\begin{eqnarray}
\dot x & = & y - g(x) + \epsilon (x^{'}-x), \\
\dot y & = & - \sigma  y - \beta x + f_1 sin( \theta)  ,\\
\dot \theta & = & \omega_1,\\
\dot {x^{'}} & = & y^{'} - g(x^{'}) + \epsilon (x-x^{'}), \\
\dot {y^{'}} & = & - \sigma  y^{'} - \beta x^{'} + f_2 sin( \theta^{'}) ,\\
\dot {\theta^{'}} & = & \omega_2,
\end{eqnarray}
\label{eqn:6}
\end{subequations}
where $g(x^{'})$ is the piecewise-linear function of the second system given by
\begin{equation}
g(x^{'}) =
\begin{cases}
bx^{'}+(a-b) & \text{if $x^{'}\ge 1$}\\
ax^{'} & \text{if $|x^{'}|\le 1$}\\
bx^{'}-(a-b) & \text{if $x^{'}\le -1$}
\end{cases}
\label{eqn:7}
\end{equation}
\begin{figure}
\begin{center}
\includegraphics[scale=0.66]{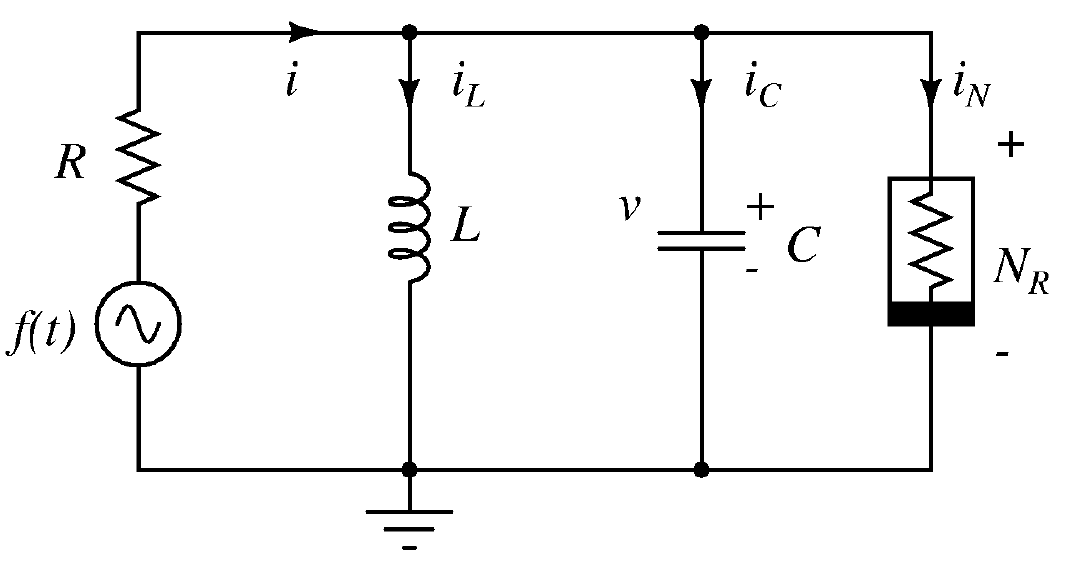}
\caption{Schematic circuit realization of the sinusoidally forced parallel LCR circuit with the nonlinear element $N_R$ connected parallel to the capacitor.}
\label{fig:4}
\end{center}
\end{figure}
A sinusoidally forced parallel LCR circuit with a nonlinear element $N_R$ connected parallel to the capacitor is as shown in Fig. \ref{fig:4}. The nonlinear element could be either a {\emph{Chua's diode}} or a {\emph{simplified nonlinear element}}. The generalized state equations for a parallel LCR circuit with a three-segmented nonlinear element can be written as 
\begin{subequations}
\begin{eqnarray} 
C {dv \over dt } & = & {1\over R} (F sin( \Omega t)- v)- i_L - g(v), \\
L {di_L \over dt } & = & v,
\end{eqnarray}
\label{eqn:8}
\end{subequations}
where the nonlinear function $g(v)$ is as given in Eq. \ref{eqn:2}. The normalized state equations are given as
\begin{subequations}
\begin{eqnarray}
\dot x & = & f sin(\theta) - x - y - g(x), \\
\dot y & = & \beta x , \\
\dot \theta & = & \omega,
\end{eqnarray}
\label{eqn:9}
\end{subequations}
The normalized circuit parameters of the circuit are different when different nonlinear elements are used. The normalized state equations of the mutually-coupled system under x-xoupling is \begin{subequations}
\begin{eqnarray}
\dot x & = & f_1 sin(\theta) - x - y - g(x) + \epsilon (x^{'}-x), \\
\dot y & = & \beta x , \\
\dot \theta & = & \omega_1,\\
\dot {x^{'}} & = & f_2 sin(\theta^{'}) - x^{'} - y^{'} - g(x^{'}) + \epsilon (x-x^{'}), \\
\dot {y^{'}} & = & \beta x^{'} , \\
\dot {\theta^{'}} & = & \omega_2,
\end{eqnarray}
\label{eqn:10}
\end{subequations}

\section{Explicit Analytical Solutions}
\label{sec:3}
In this section we present the generalized analytical solutions for the generalized state equations of the x-coupled series and parallel LCR circuits given by Eqs. \ref{eqn:6} and \ref{eqn:10}. The analytical solutions thus obtained were used to generate the synchronization dynamics of the coupled systems. 

\subsection{Series LCR circuits}

An explicit analytical solution can be presented for the normalized state equations of the series LCR circuits given in Eq. \ref{eqn:4}  for each of the piecewise linear regions $D_0$ and $D_{\pm1}$. In both the nonlinear elements, the central region has been taken as $D_0$ and the outer regions in the positive and negative voltage regime as $D_+$ and $D_-$, respectively. For both type of nonlinear elements, the fixed point corresponding to the central region is $(0,0)$. The roots ${m_{1,2}} =  \frac{-(A) \pm \sqrt{(A^{2}-4B)}} {2}$ corresponding to the $D_0$ region may be either two real values or a pair of complex conjugates. When the roots are a pair of complex conjugates, the state variables $y(t)$ and $x(t)$ are
\begin{subequations}
\begin{eqnarray}
y(t) &=& e^ {ut}(C_1 cos vt + C_2 sin vt) +E_1 + E_2 sin \omega_1 t \nonumber \\ 
&&+ E_3 cos \omega_1 t,\\
x(t) &=& \frac{1}{\beta}(-\sigma y - \dot{y} + f_1 sin \omega_1 t),
\end{eqnarray}
\label{eqn:11}
\end{subequations}
where, $u=\frac{-A}{2}$, $v=\frac{\sqrt(4B-A^{2})}{2}$ and $m_{1,2} = u \pm iv$. When the roots are two real values then the state variables are
\begin{subequations}
\begin{eqnarray}
y(t) &=& C_1 e^ {m_1 t} + C_2 e^ {m_2 t} + E_1 + E_2 \sin \omega_1 t  \nonumber \\ 
&&+ E_3 \cos \omega_1 t,\\
x(t) &=& \frac{1}{\beta}(\dot{y}), 
\end{eqnarray}
\label{eqn:12}
\end{subequations}
The fixed points $(k_1, k_2)$ corresponding to the outer regions $D_{\pm1}$ where the function $g(x)=bx \pm (a-b)$ are, $k_1$ = $\pm ((a-b) \sigma$ / $\beta+b \sigma)$, $k_2$ = $ \pm (\beta (b-a) $ / $\beta+b \sigma)$. When the roots ${m_{3,4}} =  \frac{-(A) \pm \sqrt{(A^{2}-4B)}} {2}$ are a pair of complex conjugates, the state variables are
\begin{subequations}
\begin{eqnarray}
y(t) &=& e^ {ut}(C_3 cos vt + C_4 sin vt) +E_3 sin(\omega_1 t) \nonumber \\ 
&&+ E_4 cos(\omega_1 t) {\pm} \Delta,\\
x(t) &=& \frac{1}{\beta}(\dot{y}), 
\end{eqnarray}
\label{eqn:13}
\end{subequations}
When the roots are two real values the state variables are
\begin{subequations}
\begin{eqnarray}
y(t) &=& C_3 e^ {m_3 t} + C_4 e^ {m_4 t} + E_3 \sin \omega_1 t  \nonumber \\ 
&&+ E_4 \cos \omega_1 t {\pm} \Delta,\\
x(t) &=& \frac{1}{\beta}(\dot{y}), 
\end{eqnarray}
\label{eqn:14}
\end{subequations}
where $\Delta = (b-a)$ and $+\Delta$, $-\Delta$ correspond to $D_{+1}$ and $D_{-1}$ regions respectively. \\
When the coupling parameter $\epsilon=0$, the state equations of the two systems with state variables $(x,y)$ and $(x^{'},y^{'})$ becomes identical. Hence, in the uncoupled state the explicit analytical solutions to the normalized state equations of the two systems remains the same. The solutions to the state variables $x, y$ given from Eqs. \ref{eqn:11} - \ref{eqn:14} holds good for the state variables $x^{'}, y^{'}$ except that the second system operates with a different set of initial condition. For the values of the coupling parameter $\epsilon > 0$, the x-coupled systems represented by Eq. \ref{eqn:6} are mutually bound and the dynamics of each system is controlled by the other. Now, we present explicit analytical solutions for the mutually-coupled system for coupling strengths greater than zero. Because the normalized state equations are piecewise linear, we couple each piecewise linear identical region of the first and the second system to form a new set of equation corresponding to the difference system. The difference system obtained by coupling the two set of equations given in Eq. \ref{eqn:6} are
\begin{subequations}
\begin{eqnarray}
\dot {x^{*}} & = & y^{*} - (g(x) -g(x^{'}))+ 2 \epsilon x^{*},\\
\dot {y^{*}} &= & -\sigma y^{*} - \beta x^{*} +f_1 sin(\omega_1 t) - f_2 sin(\omega_2 t),
\label{eqn:15}
\end{eqnarray}
\end{subequations}
where $x^{*}$=$(x-x^{'})$, $y^{*}$=$(y-y^{'})$ and $g(x) -g(x^{'})=g(x^{*})$ takes the values $a{x^{*}}$ or $b{x^{*}}$ depending upon the corresponding region of operation of the two systems. From these state variables $(x^{*}, y^{*})$, the state variables of the first system can be obtained as
\begin{subequations}
\begin{eqnarray}
x &=& x^{'} +  x^{*}, \\
y &=& y^{'} +  y^{*}, 
\end{eqnarray}
\label{eqn:16}
\end{subequations}
From the results of $x(t), y(t)$ from the above equations, $x^{'}(t), y^{'}(t)$ can be obtained as
\begin{subequations}
\begin{eqnarray}
x^{'} &=& x -  x^{*}, \\
y^{'} &=& y -  y^{*},
\end{eqnarray}
\label{eqn:17}
\end{subequations}
One can easily establish that a unique equilibrium point $ (x^{*}_0,y^{*}_0)$ exists for Eq. \ref{eqn:15} in each of the following three subsets
\begin{equation}
\left.
\begin{aligned}
D^{*}_{+1} & =  \{ (x^{*},y^{*})| x^{*} > 1 \}| P^{*}_+ = (0,0),\\
D^{*}_0 & =  \{ (x^{*},y^{*})|| x^{*} | < 1 \}| O^{*} = (0,0),\\
D^{*}_{-1} & =  \{ (x^{*},y^{*})|x^{*} < -1 \}| P^{*}_- = (0,0),\\
\end{aligned}
\right\}
\quad\text{}
\label{eqn:18}
\end{equation}
Since the two system differ only by their initial conditions, the origin $(0,0)$ becomes the fixed points for all the three regions of the difference system, as expected. The stability of the fixed points given in Eq. \ref{eqn:11} can be calculated from the stability matrices. In the first case, $g(x)$ and $g(x^{'})$ take the values  $a{x}$ and $a{x^{'}}$ respectively, corresponding to the central region in the $(v-i)$ characteristics of the nonlinear element, which has been taken as the $D^{*}_0$ region of the difference system. In the $D^{*}_0$ region the stability determining eigenvalues are calculated from the stability matrix
\begin{equation}
J^{*}_0 =
\begin{pmatrix}
-(a+2\epsilon) &&& 1 \\
-\beta &&& -\sigma \\
\end{pmatrix}
\label{eqn:19}
\end{equation}

In the second case, $g(x)$ and $g(x^{'})$ take the values  $bx \pm (a-b)$ and $bx^{'} \pm (a-b)$ respectively. In these regions, the stability determining eigenvalues are calculated from the stability matrix
\begin{equation}
J^{*}_{\pm} =
\begin{pmatrix}
-(b+2\epsilon) &&& 1 \\
-\beta &&& -\sigma \\
\end{pmatrix}
\label{eqn:20}
\end{equation}
The eigenvalues of the difference system are thus determined by the coupling parameter in all the three piecewise linear regions. An explicit analytical solution to the difference system given by Eq. \ref{eqn:17} can be obtained for each of the piecewise linear regions. The solutions of those equations are, $ [x^{*} (t; t_0, x^{*}_0, y^{*}_0), ~y^{*}(t; t_0, x^{*}_0, y^{*}_0)]^T$ for which the initial conditions are written as $ (t, x^{*}, y^{*}) $ $ = (t_0, x^{*}_0, y^{*}_0) $. From the solution $x^{*}(t)$ and $y^{*}(t)$ thus obtained the state variables $x(t), y(t)$ and $x^{'}(t), y^{'}(t)$  can be found using Eqs. \ref{eqn:16}. 

\subsubsection{ \bf $Region: D^{*}_0$}

In this region $g(x)$ and $g(x^{'})$ take the values  $a{x}$ and $a{x^{'}}$ respectively. Hence the normalized equations obtained from Eqs. \ref{eqn:15} are
\begin{subequations}
\begin{eqnarray}
\dot {x^{*}} & = & y^{*} - (a+2 \epsilon){x^{*}} \\
\dot {y^{*}} & = & -\sigma y^{*} - \beta x^{*}+f_1 sin(\omega_1 t) - f_2 sin(\omega_2 t) 
\end{eqnarray}
\label{eqn:21}
\end{subequations}
Differentiating Eq. \ref{eqn:21}(b) with respect to time and using Eqs. \ref{eqn:21}(a), \ref{eqn:21}(b) in the resultant equation, we obtain
\begin{eqnarray}
{\ddot y^{*}} + {A \dot y^{*}} + By^{*} = && (a+2 \epsilon) f_1~sin \omega_1 t + f_1 \omega_1~cos \omega_1 t - \nonumber \\ 
&& (a+2 \epsilon) f_2~sin \omega_2 t  - f_2 \omega_2~cos \omega_2 t
\label{eqn:22}
\end{eqnarray}
where, $ A = \sigma + a +2 \epsilon$ and $ B = \beta + \sigma(a+2 \epsilon)$. 
The roots of the Eq. \ref{eqn:22} are given by ${m_{1,2}} =  \frac{-(A) \pm \sqrt{(A^{2}-4B)}} {2}$. Since the roots ${m_{1,2}}$ depends on the coupling parameter, the orientation of the trajectories around the fixed point changes as the coupling parameter is varied. 

For $(A^{2} > 4B)$, the roots $m_1$ and $m_2$ are real and distinct. The general solution to Eq. \ref{eqn:22} can be written as
\begin{eqnarray}
y^{*}(t) = && C_1 e^ {m_1 t} + C_2 e^ {m_2 t} + E_1 sin(\omega_1 t) + E_2 cos(\omega_1 t) \nonumber \\
&& + E_3 sin(\omega_2 t) + E_4 cos(\omega_2 t)
\label{eqn:23}
\end{eqnarray}
where $C_1$ and $C_2$ are integration constants and
\begin{subequations}
\begin{eqnarray}
E_1  &=&  \frac {f_1  {\omega_1} ^2 (A-a-\epsilon) + f_1 B(a+\epsilon)}{A^2 {\omega_1} ^2 + (B-{\omega_1} ^2)^2}  \\
E_2  &=&  \frac {f_1  \omega_1 ((B-{\omega_1} ^2)-A(a+\epsilon ))}{A^2 {\omega_1} ^2 + (B-{\omega_1} ^2)^2}  \\
E_3  &=&   -\frac {f_2  {\omega_2} ^2 (A-a-\epsilon) + f_2 B(a+\epsilon)}{A^2 {\omega_2} ^2 + (B-{\omega_2} ^2)^2}  \\
E_4  &=&  -\frac {f_2  \omega_2 ((B-{\omega_2} ^2)-A(a+\epsilon ))}{A^2 {\omega_2} ^2 + (B-{\omega_2} ^2)^2}
\end{eqnarray}
\label{eqn:24}
\end{subequations}
Differentiating Eq. \ref{eqn:23} and using it in Eq. \ref{eqn:21}(b) we get,
\begin{equation}
x^{*}(t) = (\frac{1}{\beta})(\dot{y^{*}} - \sigma y^{*} + f_1 sin(\omega_1 t) - f_2 sin(\omega_2 t)),
\label{eqn:25}
\end{equation}
The constants $C_1$ and $C_2$ are given as
\begin{eqnarray}
C_1 =  &&\frac{e^ {- m_1 t_0}} {m_1 - m_2} \{ (-\sigma{ y^{*}_0}-\beta{ x^{*}_0}-m_2{ y^{*}_0}) \nonumber\\
&& - (\omega_1 E_1 - m_2 E_2) cos \omega_1 t_0 \nonumber \\
&& + ( f_1+ \omega_1 E_2 + m_2 E_1) sin \omega_1 t_0  \nonumber \\
&&  -(\omega_2 E_3 - m_2 E_4) cos \omega_2 t_0 \nonumber \\
&&  + (\omega_2 E_4 + m_2 E_3 - f_2) sin \omega_2 t_0 \} \nonumber
\end{eqnarray}
\begin{eqnarray}
C_2 =  && \frac{e^ {- m_2 t_0}} {m_2 - m_1} \{ (-\sigma{ y^{*}_0}-\beta{ x^{*}_0}-m_1{ y^{*}_0}) \nonumber \\
&& - (\omega_1 E_1 - m_1 E_2) cos \omega_1 t_0 \nonumber \\
&& + ( f_1+ \omega_1 E_2 + m_1 E_1) sin \omega_1 t_0 \nonumber \\
&& -(\omega_2 E_3 - m_1 E_4) cos \omega_2 t_0 \nonumber \\
&& + (\omega_2 E_4 + m_1 E_3 - f_2) sin \omega_2 t_0\} \nonumber 
\end{eqnarray}
From the results obtained for $y^{*}(t), x^{*}(t)$ in Eqs. \ref{eqn:23} and \ref{eqn:25}, $y(t),x(t)$ can be obtained from Eqs. \ref{eqn:16} and using their values so obtained, $x^{'}(t)$ and $y^{'}(t)$ can be obtained from Eqs. \ref{eqn:17}. Hence, the state variables of the first and the second system depends on each other at every instant and evolves with time.\\
For $(A^{2} < 4B)$, the roots $m_1$ and $m_2$ are a pair of complex conjugates given as $m_{1,2} = u \pm iv$, with $u=\frac{-A}{2}$ and $v=\frac{\sqrt(4B-A^{2})}{2}$.
The general solution to Eq. \ref{eqn:22} can be written as
\begin{eqnarray}
y^{*}(t) =  && e^ {ut} (C_3 cosvt+ C_4 sinvt)+ E_5 sin \omega_1 t + E_6 cos \omega_1 t \nonumber \\
&& + E_7 sin \omega_2 t + E_8 cos \omega_2 t 
\label{eqn:26}
\end{eqnarray}
where the constants $E_5,E_6,E_7,E_8$ are the same as that of $E_1,E_2,E_3,E_4$ given in Eqs. \ref{eqn:24}, respectively.
Differentiating Eq. \ref{eqn:26} and using it in Eq. \ref{eqn:20}(b) we get,
\begin{equation}
x^{*}(t) = (\frac{1}{\beta})(-\dot{y^{*}} - \sigma y^{*}+ f_1 sin(\omega_1 t) - f_2 sin(\omega_2 t))
\label{eqn:27}
\end{equation}
The constants $C_3$ and $C_4$ are given as,
\begin{eqnarray}
C_3 = && \frac{e^ {- u t_0}} {v} \{((\sigma { y^{*}_0} +\beta { x^{*}_0}+u { y^{*}_0})sinvt_0 + v { y^{*}_0}) cosvt_0 \nonumber \\
&& +((\omega_1 E_5 - u E_6) sinvt_0 - vE_6 cosvt_0)cos \omega_1 t_0 \nonumber \\
&& - ((f_1+\omega_1 E_6 +u E_5) sinvt_0+v E_5 cosvt_0) sin \omega_1 t_0 \nonumber \\
&& + ((\omega_2 E_7 - u E_8) sinvt_0 - vE_8 cosvt_0)cos \omega_2 t_0  \nonumber \\
&&  - ((\omega_2 E_8 + u E_7-f_2) sinvt_0 + vE_7 cosvt_0) sin \omega_2 t_0 \}  \nonumber
\end{eqnarray}
\begin{eqnarray}	 
C_4 = && \frac{e^ {- u t_0}} {v} \{((-\sigma { y^{*}_0} -\beta { x^{*}_0} - u { y^{*}_0})cos vt_0 + v { y^{*}_0}) sin vt_0 \nonumber \\
&& - ((\omega_1 E_5 - u E_6) cos vt_0 + vE_6 sin vt_0)cos \omega_1 t_0 \nonumber \\
&&  + ((f_1+\omega_1 E_6 +u E_5) cos vt_0 - v E_5 sin vt_0) sin \omega_1 t_0 \nonumber \\
&& - ((\omega_2 E_7 - u E_8) cos vt_0 + vE_8 sin vt_0)cos \omega_2 t_0  \nonumber \\
&& - ((\omega_2 E_8 + u E_7 - f_2) cos vt_0 - vE_7 sin vt_0) sin \omega_2 t_0 \}  \nonumber 
\end{eqnarray}\\
From the results obtained for $y^{*}(t), x^{*}(t)$ in Eqs. \ref{eqn:26} and \ref{eqn:27}, $y(t),x(t)$ can be obtained from Eqs. \ref{eqn:16} and using their values so obtained, $x^{'}(t)$ and $y^{'}(t)$ can be obtained from Eqs. \ref{eqn:17}.

\subsubsection{ \bf $Region: D^{*}_{\pm1}$}

In the $D^{*}_{\pm1}$ region, $g(x)$, $g(x^{'})$ take the values $bx \pm (a-b)$, $bx^{'} \pm (a-b)$, respectively and hence $g(x^{*}) = bx^{*}$. The normalized state equations can be written as
\begin{subequations}
\begin{eqnarray}
\dot {x^{*}} & = & y^{*} - (b+2 \epsilon){x^{*}} \\
\dot {y^{*}} & = & -\sigma y^{*} - \beta x^{*}+f_1 sin(\omega_1 t) - f_2 sin(\omega_2 t) 
\end{eqnarray}
\label{eqn:28}
\end{subequations}

Differentiating Eq. \ref{eqn:28}(b) with respect to time and using Eqs. \ref{eqn:28}(a), (b) in the resultant equation, we obtain 
\begin{eqnarray}
{\ddot y^{*}} + {C \dot y^{*}} + B y^{*} = && (b+2 \epsilon) f_1~sin \omega_1 t + f_1 \omega_1~cos \omega_1 t - \nonumber \\ 
&& (b+2 \epsilon) f_2~sin \omega_2 t  - f_2 \omega_2~cos \omega_2 t
\label{eqn:29}
\end{eqnarray}
where, $ C = \sigma + b +2 \epsilon$ and $ D = \beta + \sigma(b+2 \epsilon)$. For $C^{2} > 4D$ the roots are a real and distinct and are given by $m_{3,4} = \frac{-C \pm \sqrt{(C^{2}-4D)}} {2}$. Hence the general solution to Eq. \ref{eqn:29} can be written as
\begin{eqnarray}
y^{*}(t) = && C_1 e^ {m_3 t} + C_2 e^ {m_4 t} + E_1 sin(\omega_1 t) + E_2 cos(\omega_1 t) \nonumber \\
&& + E_3 sin(\omega_2 t) + E_4 cos(\omega_2 t)
\label{eqn:30}
\end{eqnarray}
where $C_1, C_2, E_1, E_2, E_3, E_4$ are the same as their counterparts in the $D^{*}_0$ region except that the constants $A$ and $B$ are replaced with the constants $C$ and $D$. Differentiating Eq. \ref{eqn:30} and using it in Eq. \ref{eqn:28}(b) we get
\begin{equation}
x^{*}(t) = (\frac{1}{\beta})(-\dot{y^{*}} - \sigma y^{*}+ f_1 sin(\omega_1 t) - f_2 sin(\omega_2 t))
\label{eqn:31}
\end{equation}
From the results obtained for $y^{*}(t), x^{*}(t)$ in Eqs. \ref{eqn:29} and \ref{eqn:31}, $y(t),x(t)$ can be obtained from Eqs. \ref{eqn:16} and using their values so obtained, $x^{'}(t)$ and $y^{'}(t)$ can be obtained from Eqs. \ref{eqn:17}.\\
For $C^2 < 4D$, the roots are a pair of complex conjugates given as $m_{3,4} = u \pm iv$, with $u=\frac{-C}{2}$ and $v=\frac{\sqrt(4D-C^{2})}{2}$. Hence the general solution to Eq. \ref{eqn:29} can be written as
\begin{eqnarray}
y^{*}(t)  =&&  e^ {ut} (C_3 cosvt+ C_4 sinvt)+ E_5 sin \omega_1 t + E_6 cos \omega_1 t  \nonumber \\
&& + E_7 sin \omega_2 t + E_8 cos \omega_2 t 
\label{eqn:32}
\end{eqnarray}
The constants $C_3,C_4, E_5, E_6, E_7, E_8$ are the same as given in $D_0$ region except that the constants $A$ and $B$ are replaced with the constants $C$ and $D$.
Differentiating Eq. \ref{eqn:32} and using it in Eq. \ref{eqn:28}(b) we get
\begin{equation}
x^{*}(t) = (\frac{1}{\beta})(-\dot{y^{*}} - \sigma y^{*}+ f_1 sin(\omega_1 t) - f_2 sin(\omega_2 t))
\label{eqn:33}
\end{equation}
From the results obtained for $y^{*}(t), x^{*}(t)$ in Eqs. \ref{eqn:32} and \ref{eqn:33}, $y(t),x(t)$ can be obtained from Eqs. \ref{eqn:16} and using their values so obtained, $x^{'}(t)$ and $y^{'}(t)$ can be obtained from Eqs. \ref{eqn:17}.\\
\begin{figure}
\begin{center}
\includegraphics[scale=0.33]{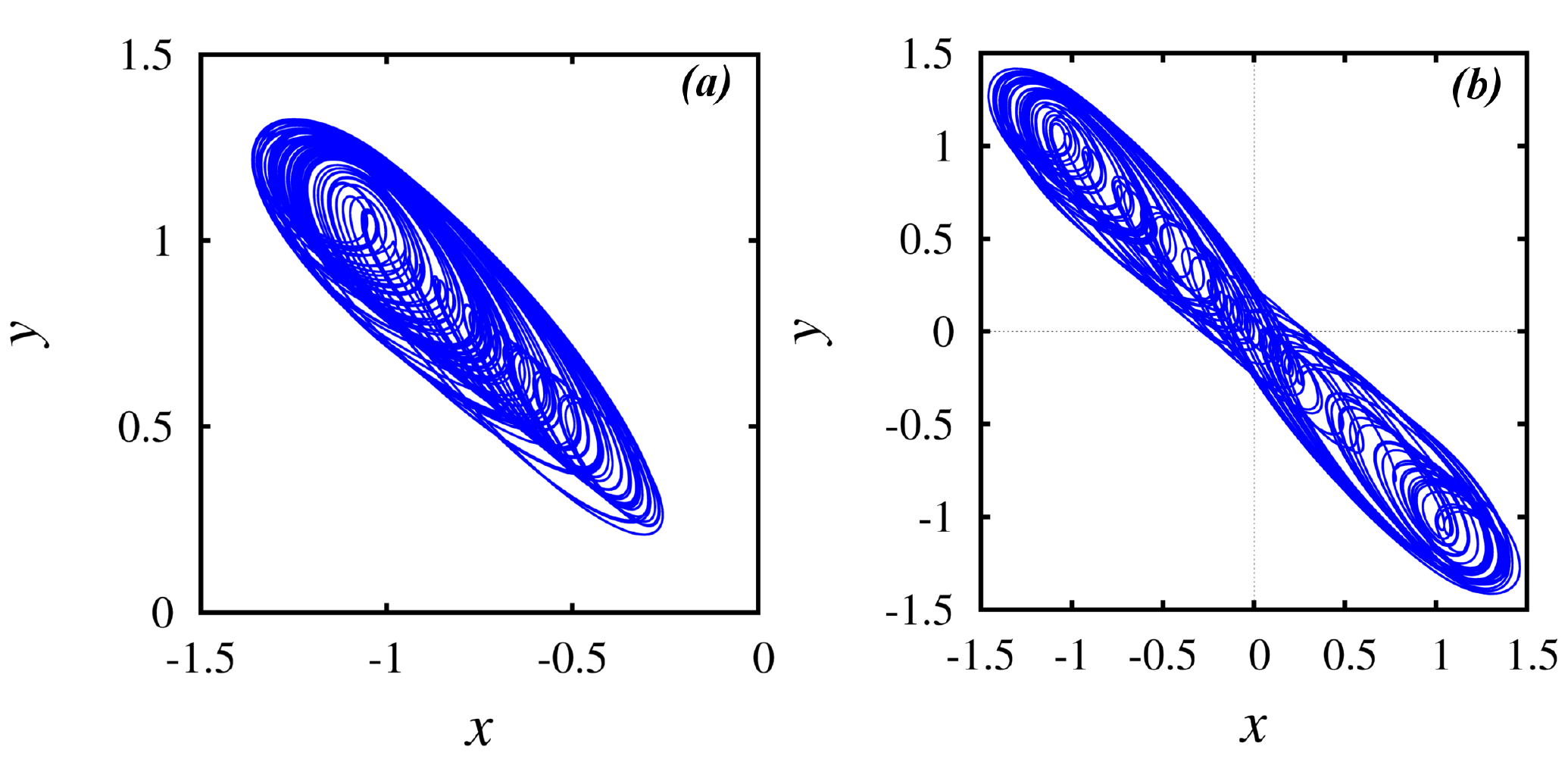}
\caption{(Color online) Analytically obtained (a) One-band and (b) Double-band chaotic attractors of the {\emph{Murali-Lakshmanan-Chua}} circuit at the control parameter values $f=0.1$ and $f=0.14$, respectively.}
\label{fig:5}
\end{center}
\end{figure}
\begin{figure}
\begin{center}
\includegraphics[scale=0.33]{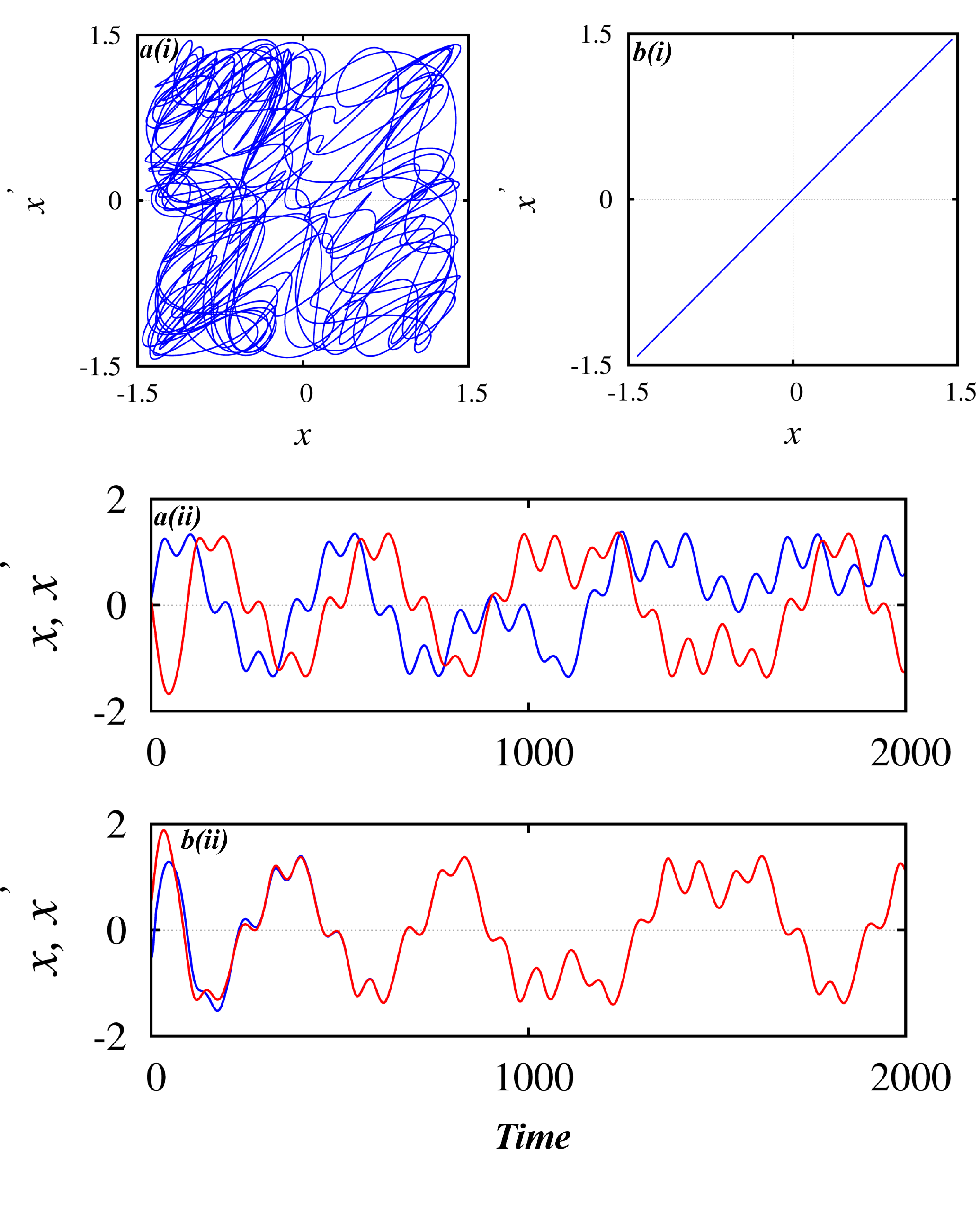}
\caption{(Color online) Analytically obtained synchronization dynamics of the {\emph{MLC}} circuit for coupled double-band chaotic attractors; a(i) Unsynchronized states in $(x-x^{'})$ phase plane and (ii) the corresponding time series of the signals $x$ (blue line) and $x^{'}$ (red line) for $\epsilon = 0$; b(i) The synchronized state in the $(x-x^{'})$ phase plane and (ii) the corresponding time series of the signals $x$ (blue line) and $x^{'}$ (red line) for $\epsilon = 0.1$.}
\label{fig:6}
\end{center}
\end{figure}
%
Now let us briefly explain how the solution can be generated in the $(x^{'}-y^{'})$ phase space. The analytical solutions obtained above can be used to simulate the trajectories of the state variables $y^{*}(t)$ and $x^{*}(t)$. With the {\emph{time (t)}} being considered as the independent variable, the state variables evolves within each piecewise-linear region depending upon its initial values. If we start with the initial conditions $x^{*}(t=0) = x^{*}_0,~y^{*}(t=0) = y^{*}_0$ in the $D^{*}_0$ region at time $t=0$, the arbitrary constants $C_1$ and $C_2$ get fixed. Thus $x^{*}(t)$ evolves as given by Eq. \ref{eqn:27} up to either $t=T_1$, when $x^{*}(T_1)=1$ or $t = T^{'}_1$ when $x^{*}(T^{'}_1) = -1$. The next region of operation $(D^{*}_{+1}$ or $D^{*}_{-1})$ thus depends upon the value of $x^{*}$ in the $D^{*}_{0}$ region at that instant of time. As the trajectory enters into the next region of interest, the arbitrary constant corresponding to that region could be evaluated, with the initial conditions to that region being either ($x^{*}_{0}(T_1),~y^{*}_{0}(T_1)$) or  ($x^{*}_{0}(T^{'}_1),~y^{*}_{0}(T^{'}_1)$). During each region of operation, the state variables of the response system evolves as $x^{'}(t) = x(t) - x^{*}(t)$ and $y^{'}(t) = y(t) - y^{*}(t)$, respectively. The procedure can be continued for each successive crossing. In this way, the explicit solutions can be obtained in each of the regions $D^{'}_0$, $D^{'}_{\pm1}$ of the response system. The solution obtained in each region has been matched across the boundaries and used to generate the dynamics of the response system.  \\ 
From the analytical solutions obtained above, the synchronization dynamics of mutually-coupled series LCR circuits, under x-coupling can be studied. First we present the complete synchronization phenomenon exhibited by the coupled-circuits with a {\emph{Chua's diode}} as the nonlinear element, called the {\emph{MLC}} circuit. The MLC circuit exhibits two prominent chaotic attractors at the amplitude of the external force $f=0.1$ and $f=0.14$, respectively as shown in Fig. \ref{fig:6}. The mutually-coupled systems which are initially unsynchronized for the coupling strength $\epsilon = 0$, becomes completely synchronized at higher values of the coupling strength. The trajectories of the of the two systems enters into a stable synchronization manifold and exists thereafter. Fig. \ref{fig:7} shows the unsynchronized state for $\epsilon =0$, the completely synchronized state for $\epsilon = 0.1$ and their corresponding time-series plots of the state variable $x, x^{'}$, respectively.\\
A similar study can be made for the coupled systems each with a {\emph{simplified nonlinear element}}. The forced series LCR circuit with a simplified nonlinear element exhibits a prominent chaotic attractor at the value of the forcing amplitude $f = 0.31$ as shown in Fig. \ref{fig:7}. The synchronization dynamics of the x-coupled system obtained from the analytical solutions is shown in Fig. \ref{fig:8}. The unsynchronized state for $\epsilon = 0$ and the complete synchronization state for $\epsilon = 0.12$ along with the time-series of the signals indicating synchronization are shown in Fig. \ref{fig:8}. 

\subsection{Parallel LCR circuits}

An explicit analytical solution to the x-coupled, normalized state equations of the mutually-coupled parallel LCR circuits given in Eq. \ref{eqn:10} can be obtained in a similar way as that of the series LCR circuits. The explicit analytical solutions to the state equations of the uncoupled system given in Eq. \ref{eqn:9} can be summarized as follows.\\
In the $D_0$ region, $g(x)=ax$ with $(0,0)$ as the fixed point. The roots ${m_{1,2}} =  \frac{-(A) \pm \sqrt{(A^{2}-4B)}} {2}$ are either two real values or a pair of complex conjugates. When the roots are a pair of complex conjugates, the state variables $y(t)$ and $x(t)$ are
\begin{subequations}
\begin{eqnarray}
y(t) &=& e^ {ut}(C_1 cos vt + C_2 sin vt) +E_1 sin \omega_1 t \nonumber \\ 
&&+ E_2 cos \omega_1 t,\\
x(t) &=& \frac{1}{\beta}(\dot{y}), 
\end{eqnarray}
\label{eqn:34}
\end{subequations}
where, $u=\frac{-A}{2}$ and $v=\frac{\sqrt(4B-A^{2})}{2}$. When the roots are two real values then the state variables are
\begin{subequations}
\begin{eqnarray}
y(t) &=& C_1 e^ {m_1 t} + C_2 e^ {m_2 t} + E_1 + E_2 \sin \omega_1 t  \nonumber \\ 
&&+ E_3 \cos \omega_1 t,\\
x(t) &=& \frac{1}{\beta}(\dot{y}), 
\end{eqnarray}
\label{eqn:35}
\end{subequations}
In outer regions $D_{\pm 1}$, we have $g(x)=bx \pm (a-b)$ with $k_1=0$, $k_2={\pm}(a-b)$ as the fixed point. When the roots are a pair of complex conjugates given by ${m_{3,4}} =  \frac{-(A) \pm \sqrt{(A^{2}-4B)}} {2}$, the state variables $y(t)$ and $x(t)$ are
\begin{subequations}
\begin{eqnarray}
y(t) &=& e^ {ut}(C_3 cos vt + C_4 sin vt) +E_3 sin(\omega_1 t) \nonumber \\ 
&&+ E_4 cos(\omega_1 t) {\pm} \Delta,\\
x(t) &=& \frac{1}{\beta}(\dot{y}), 
\end{eqnarray}
\label{eqn:36}
\end{subequations}
When the roots are two real values then the state variables are
\begin{subequations}
\begin{eqnarray}
y(t) &=& C_3 e^ {m_3 t} + C_4 e^ {m_4 t} + E_3 \sin \omega_1 t  \nonumber \\ 
&&+ E_4 \cos \omega_1 t {\pm} \Delta,\\
x(t) &=& \frac{1}{\beta}(\dot{y}), 
\end{eqnarray}
\label{eqn:37}
\end{subequations}
where $\Delta = (b-a)$ and $+\Delta$, $-\Delta$ corresponds to $D_{+1}$ and $D_{-1}$ regions respectively. The solutions given above are also valid for the state variables of the second system $(x^{'}, y^{'})$ except that the two systems operate with different set of initial conditions.\\
The difference system obtained form Eq. \ref{eqn:10} can be written as
\begin{subequations}
\begin{eqnarray}
\dot {x^{*}} &=& f_1 sin(\omega_1 t) - f_2 sin(\omega_2 t) - x^{*} - y^{*}  \nonumber \\ 
&&- g(x^{*})- 2 \epsilon x^{*},\\
\dot {y^{*}} &=& \beta x^{*}, 
\end{eqnarray}
\label{eqn:38}
\end{subequations}
where $x^{*}$=$(x-x^{'})$, $y^{*}$=$(y-y^{'})$ and $g(x^{*}) = g(x) - g(x^{'})$ takes the values $a{x^{*}}$ or $b{x^{*}}$ depending upon the region of operation of the drive and response system. From the new set of state variables $x^{*}(t),~y^{*}(t)$, the state variables $x(t),~y(t)$ of the first system can be obtained as
\begin{subequations}
\begin{eqnarray}
x &=& x^{'} +  x^{*}, \\
y &=& y^{'} +  y^{*},
\end{eqnarray}
\label{eqn:39}
\end{subequations}
From the results of $x(t), y(t)$ from the above equations, $x^{'}(t), y^{'}(t)$ can be obtained as
\begin{subequations}
\begin{eqnarray}
x^{'} &=& x -  x^{*}, \\
y^{'} &=& y -  y^{*},
\end{eqnarray}
\label{eqn:40}
\end{subequations}
The equilibrium points corresponding to all the three regions of the difference system is as given in Eq. \ref{eqn:18}.
\begin{equation}
\left.
\begin{aligned}
D^{*}_{+1} & =  \{ (x^{*},y^{*})| x^{*} > 1 \} P^{*}_+ = (0,0),\\
D^{*}_0 & =  \{ (x^{*},y^{*})|| x^{*} | < 1 \}| O^{*} = (0,0),\\
D^{*}_{-1} & =  \{ (x^{*},y^{*})|x^{*} < -1 \}| P^{*}_- = (0,0),\\
\end{aligned}
\right\}
\quad\text{}
\label{eqn:41}
\end{equation}
The stability of the fixed points $i.e.$ the origin, can be calculated from the stability matrices. In the $D^{*}_0$ region the stability determining eigenvalues are determined from the stability matrix
\begin{equation}
J^{*}_{0} =
\begin{pmatrix}
-(a+2 \epsilon+1) &&& -1 \\
\beta &&& 0 \\
\end{pmatrix},
\label{eqn:42}
\end{equation}
In the $D^{*}_{\pm1}$ regions, $g(x)$ and $g(x^{'})$ take the values  $bx \pm (a-b)$ and $bx^{'} \pm (a-b)$ respectively. The stability determining eigenvalues in these regions can be determined from the stability matrix
\begin{equation}
J^{*}_{\pm} =
\begin{pmatrix}
-(b+2 \epsilon+1) &&& -1 \\
\beta &&& 0 \\
\end{pmatrix}
\label{eqn:43}
\end{equation}
Now we present explicit analytical solutions for the normalized state equations of the mutually-coupled system for coupling strengths $\epsilon > 0$. 

\subsubsection{ \bf $Region: D^{*}_0$}

In this region $g(x)$ and $g(x^{'})$ takes the values  $a{x}$ and $a{x^{'}}$ respectively. Hence the normalized equations obtained from Eqs. \ref{eqn:38} are
\begin{subequations}
\begin{eqnarray}
\dot {x^{*}} & = & - (a+2 \epsilon+1)x^{*} - y^{*}+f_1 sin(\omega_1 t) \nonumber \\ 
&&- f_2 sin(\omega_2 t), \\
\dot {y^{*}} & = & \beta x^{*},
\end{eqnarray}
\label{eqn:44}
\end{subequations}
Differentiating Eq. \ref{eqn:44}(b) with respect to time and using Eqs. \ref{eqn:44}(a), \ref{eqn:44}(b) in the resultant equation, we obtain
\begin{equation}
{\ddot y^{*}} + {A \dot y^{*}} + By^{*} = f_1 sin(\omega_1 t) - f_2 sin(\omega_2 t),
\label{eqn:45}
\end{equation}
where, $ A =  a + 2 \epsilon +1$ and  $B = \beta$.
The roots of the Eq. \ref{eqn:45} is given by ${m_{1,2}} =  \frac{-(A) \pm \sqrt{(A^{2}-4B)}} {2}$. When $(A^{2} < 4B)$, the roots $m_1$ and $m_2$ are a pair of complex conjugates given as $m_{1,2} = u \pm iv$, with $u=\frac{-A}{2}$ and $v=\frac{\sqrt(4B-A^{2})}{2}$. Hence, the general solution to Eq. \ref{eqn:45} can be written as
\begin{eqnarray}
y^{*}(t) &=&  e^ {ut} (C_1 cosvt+ C_2 sinvt)+ E_1 sin \omega_1 t + E_2 cos \omega_1 t  \nonumber \\
&&+ E_3 sin \omega_2 t + E_4 cos \omega_2 t,
\label{eqn:46}
\end{eqnarray}
where $C_1$ and $C_2$ are integration constants and
\begin{subequations}
\begin{eqnarray}
E_1  &=&  \frac {f_1 \beta (B-{\omega_1}^2)}{A^2 {\omega_1} ^2 + (B-{\omega_1} ^2)^2},  \\
E_2  &=&  \frac {-A \omega_1 f_1 \beta}{A^2 {\omega_1} ^2 + (B-{\omega_1} ^2)^2},  \\
E_3  &=&   \frac {-f_2 \beta (B-{\omega_2}^2)}{A^2 {\omega_2} ^2 + (B-{\omega_2} ^2)^2},  \\
E_4  &=&  \frac {A \omega_2 f_2 \beta}{A^2 {\omega_2} ^2 + (B-{\omega_2} ^2)^2}, 
\end{eqnarray}
\label{eqn:47}
\end{subequations}
Differentiating Eq. \ref{eqn:46} and using it in Eq. \ref{eqn:44}(b) we get
\begin{equation}
x^{*}(t) = \frac{1}{\beta}(\dot{y^{*}}),
\label{eqn:48}
\end{equation}
The constants $C_1$ and $C_2$ are
\begin{eqnarray}
C_1 &=&  \frac{e^ {- u t_0}} {v} \{(v { y^{*}_0} cos vt_0- (\beta { x^{*}_0}-u { x^{*}_0}) sin vt_0) \nonumber \\ 
            &&+((\omega_1 E_1 - u E_2) sinvt_0 - vE_2 cosvt_0)cos \omega_1 t_0 \nonumber \\
            &&- ((\omega_1 E_2 +u E_1) sinvt_0+v E_1 cosvt_0) sin \omega_1 t_0 \nonumber \\ 
            &&+ ((\omega_2 E_3 - u E_4) sinvt_0 - vE_4 cosvt_0)cos \omega_2 t_0  \nonumber \\
	 && - ((\omega_2 E_4 + u E_3) sinvt_0 + vE_3 cosvt_0) sin \omega_2 t_0 \},  \nonumber \\
C_2 &=&  \frac{e^ {- u t_0}} {v} \{((\beta { x^{*}_0}-u { x^{*}_0}) cos vt_0+v { y^{*}_0} sin vt_0)  \nonumber \\
           &&-((\omega_1 E_1 - u E_2) cos vt_0 + v E_2 sin vt_0)cos \omega_1 t_0 \nonumber \\
           &&+ ((\omega_1 E_2 +u E_1) cos vt_0 - v E_1 sin vt_0) sin \omega_1 t_0  \nonumber \\
           && - ((\omega_2 E_3 - u E_4) cos vt_0 + vE_4 sin vt_0)cos \omega_2 t_0  \nonumber \\
	&& + ((\omega_2 E_4 + u E_3) cos vt_0 - vE_3 sin vt_0) sin \omega_2 t_0 \}, \nonumber 
\end{eqnarray}
From the results obtained for $y^{*}(t), x^{*}(t)$ in Eqs. \ref{eqn:46} and \ref{eqn:48}, $y(t),x(t)$ can be obtained from Eqs. \ref{eqn:39} and using their values so obtained, $x^{'}(t)$ and $y^{'}(t)$ can be obtained from Eqs. \ref{eqn:40}.\\
When $(A^{2} > 4B)$, the roots $m_1$ and $m_2$ are real and distinct. The general solution to Eq. \ref{eqn:45} can be written as
\begin{eqnarray}
y^{*}(t) &=& C_1 e^ {m_1 t} + C_2 e^ {m_2 t} + E_1 sin(\omega_1 t) + E_2 cos(\omega_1 t)  \nonumber \\ 
&& + E_3 sin(\omega_2 t) + E_4 cos(\omega_2 t),
\label{eqn:49}
\end{eqnarray}
where $C_1$ and $C_2$ are the integration constants and the constants $E_1,E_2,E_3,E_4$ are the same as  Eq. \ref{eqn:47}.
Differentiating Eq .\ref{eqn:49} and using it in Eq. \ref{eqn:44}(b) we get
\begin{equation}
x^{*}(t) = \frac{1}{\beta}(\dot{y^{*}}),
\label{eqn:50}
\end{equation}
The constants $C_1$ and $C_2$ are
\begin{eqnarray}
C_1 &=&  \frac{e^ {- m_2 t_0}} {m_2 - m_1} \{ (\beta{ x^{*}_0} - m_1{ y^{*}_0}) + ( m_1 E_2 - \omega_1 E_1) cos \omega_1 t_0  \nonumber \\
            && + (\omega_1 E_2 + m_1 E_1) sin \omega_1 t_0 +(m_2 E_4 - \omega_2 E_3) cos \omega_2 t_0  \nonumber \\
            &&+ (\omega_2 E_4 + m_2 E_3 ) sin \omega_2 t_0 \}, \nonumber \\
C_2 &=&  \frac{e^ {- m_1 t_0}} {m_1 - m_2} \{ (\beta{ x^{*}_0} - m_2{ y^{*}_0}) + ( m_2 E_2 - \omega_1 E_1) cos \omega_1 t_0  \nonumber \\
            &&+ (\omega_1 E_2 + m_2 E_1) sin \omega_1 t_0 +(m_1 E_4 - \omega_2 E_3) cos \omega_2 t_0  \nonumber \\
            &&+ (\omega_2 E_4 + m_1 E_3 ) sin \omega_2 t_0 \}, \nonumber
\end{eqnarray} 
From the results obtained for $y^{*}(t), x^{*}(t)$ in Eqs. \ref{eqn:49} and \ref{eqn:50}, $y(t),x(t)$ can be obtained from Eqs. \ref{eqn:39} and using their values so obtained, $x^{'}(t)$ and $y^{'}(t)$ can be obtained from Eqs. \ref{eqn:40}.\\

\subsubsection{\bf $Region: D^{*}_{\pm1}$}

In this region $g(x)$ and $g(x^{'})$ takes the values  $b{x}\pm (a-b)$ and $b{x^{'}} \pm (a-b)$ respectively. Hence the normalized state equations obtained from Eq. \ref{eqn:38} are
\begin{subequations}
\begin{eqnarray}
\dot {x^{*}} & = & - (b+2 \epsilon+1)x^{*} - y^{*}+f_1 sin(\omega_1 t)  \nonumber \\
&& - f_2 sin(\omega_2 t), \\
\dot {y^{*}} & = & \beta x^{*},
\end{eqnarray}
\label{eqn:51}
\end{subequations}
Differentiating Eq. \ref{eqn:51}(b) with respect to time and using Eqs. \ref{eqn:51}(a), \ref{eqn:51}(b) in the resultant equation, we obtain
\begin{equation}
{\ddot y^{*}} + {C \dot y^{*}} + Dy^{*} = f_1 sin(\omega_1 t) - f_2 sin(\omega_2 t),
\label{eqn:52}
\end{equation}
where, $ C =  b + 2 \epsilon +1$ and  $D = \beta$.
The roots of the Eq. \ref{eqn:52} is given by ${m_{3,4}} =  \frac{-(C) \pm \sqrt{(C^{2}-4D)}} {2}$. When $(C^{2} < 4D)$, the roots $m_3$ and $m_4$ are a pair of complex conjugates given as $m_{3,4} = u \pm iv$, with $u=\frac{-C}{2}$ and $v=\frac{\sqrt(4D-C^{2})}{2}$. The state variables $y^{*}(t), x^{*}(t)$ can be written as
\begin{subequations}
\begin{eqnarray}
y^{*}(t) &=&  e^ {ut} (C_3 cosvt+ C_4 sinvt)+ E_5 sin \omega_1 t + E_6 cos \omega_1 t  \nonumber \\
&& + E_7 sin \omega_2 t + E_8 cos \omega_2 t,\\
x^{*}(t) &=& \frac{1}{\beta}(\dot{y^{*}}),
\end{eqnarray}
\label{eqn:53}
\end{subequations}
The constants $C_3, C_4$ and $E_5, E_6, E_7, E_8$ are the same  as the constants $C_1, C_2$ $E_1, E_2, E_3, E_4$ in $D^{*}_{0}$ region except that the constants $A,~B$ are replaced with $C,~D$ respectively.\\ 
When $(C^{2} > 4D)$, the roots $m_3$ and $m_4$ are real and distinct. The state variables  $y^{*}(t), x^{*}(t)$ can be written as 
\begin{subequations}
\begin{eqnarray}
y^{*}(t) &=& C_3 e^ {m_3 t} + C_4 e^ {m_4 t} + E_5 sin(\omega_1 t) + E_6 cos(\omega_1 t)  \nonumber \\ 
&&+ E_7 sin(\omega_2 t) + E_8 cos(\omega_2 t),\\
x^{*}(t) &=& \frac{1}{\beta}(\dot{y^{*}}).
\end{eqnarray}
\label{eqn:54}
\end{subequations}
The constants $C_3, C_4$ and $E_5,E_6, E_7, E_8$ are the same  as the constants $C_1, C_2$ $E_1, E_2, E_3, E_4$ in $D^{*}_{0}$ region except that the constants $A,~B$ are replaced with $C,~D$ respectively.\\ 
Now let us briefly explain how the solution can be generated in the $(x^{'}-y^{'})$ phase space. The analytical solutions obtained above can be used to simulate the trajectories of the state variables $y^{*}(t)$ and $x^{*}(t)$. With the {\emph{time (t)}} being considered as the independent variable, the state variables evolves within each piecewise-linear region depending upon its initial values. If we start with the initial conditions $x^{*}(t=0) = x^{*}_0,~y^{*}(t=0) = y^{*}_0$ in the $D^{*}_0$ region at time $t=0$, the arbitrary constants $C_1$ and $C_2$ get fixed. Thus $x^{*}(t)$ evolves as given by Eq. \ref{eqn:50} up to either $t=T_1$, when $x^{*}(T_1)=1$ or $t = T^{'}_1$ when $x^{*}(T^{'}_1) = -1$. The next region of operation $(D^{*}_{+1}$ or $D^{*}_{-1})$ thus depends upon the value of $x^{*}$ in the $D^{*}_{0}$ region at that instant of time. As the trajectory enters into the next region of interest, the arbitrary constant corresponding to that region could be evaluated, with the initial conditions to that region being either ($x^{*}_{0}(T_1),~y^{*}_{0}(T_1)$) or  ($x^{*}_{0}(T^{'}_1),~y^{*}_{0}(T^{'}_1)$). During each region of operation, the state variables of the response system evolves as $x^{'}(t) = x(t) - x^{*}(t)$ and $y^{'}(t) = y(t) - y^{*}(t)$, respectively. The procedure can be continued for each successive crossing. In this way, the explicit solutions can be obtained in each of the regions $D^{'}_0$, $D^{'}_{\pm1}$ of the response system. The solution obtained in each region has been matched across the boundaries and used to generate the dynamics of the response system.  \\ 

\section{Mutually-coupled series LCR circuits: Numerical and Experimental results}
\label{sec:4}
In this section we present numerical and experimental results for the synchronization dynamics observed in mutually-coupled series LCR circuits with two different nonlinear elements. Numerical studied are performed for the $x \leftrightarrow x^{'}$ coupling of the state variables of the systems represented by Eqs. \ref{eqn:6}. Linear stability analysis indicating the bifurcation of fixed points in each piecewise linear regions of the difference system is presented. The stability of synchronization in each circuit systems are studied using the {\emph{Master Stability Function}}. Finally, experimental results are presented to confirm the analytical and numerical simulation results.

\subsection{{\emph{Murali-Lakshmanan-Chua}} Circuits}
\label{subsubsec:1}

A sinusoidally forced series LCR circuit with the {\emph{Chua's diode}} as the nonlinear element called as the {\emph{Murali-Lakshmanan-Chua}} circuit was introduced by Murali {\emph{et al} \cite{Murali1994}. The normalized circuit parameters take the values  $\beta=1,~\nu=0.015,~\omega=0.72,~a=-1.02,$ and $b=-0.55$. The circuit exhibits a period doubling sequence to chaos and has two prominent chaotic attractors, the one-band and the double-band chaotic attractors, at the control parameter values $f=0.1$ and $f=0.14$, respectively. \\
A linear stability analysis is carried out to analyze the stability of the fixed points of the difference system in their corresponding synchronization manifold. From Eq. \ref{eqn:18} we observe that the {\emph{origin (0,0)}} acts the fixed point of the difference system for all the three regions $D^{*}_{0}$ and $D^{*}_{\pm1}$. The Jacobian matrix of the difference system obtained from the state equations of the coupled system for the $x \leftrightarrow x^{'}$ coupling in the $D^{*}_{0}$ and $D^{*}_{\pm1}$regions can be written as 
\begin{equation}
J^{*}_{0, \pm1} =
\begin{pmatrix}
-2\epsilon-
\begin{cases}
a, \text{$|x| < 1$}\\
b, \text{$|x| > 1$}\\
\end{cases}
 && 1 \\
-\beta && -\sigma \\
\end{pmatrix}, 
\label{eqn:55}
\end{equation}
In the $D^{*}_{0}$ region, the {\emph{origin}} transforms into a {\emph{stable spiral}} from a {\emph{saddle}} through the evolution of a {\emph{stable node}}. The {\emph{origin}} is a saddle for $\epsilon < 0.01739$ and is a stable node for $0.01739 \le \epsilon \le 0.01749$ and becomes a stable spiral for $\epsilon > 0.01749$. Fig.~\ref{fig:13} shows the behavior of the origin for different values of the coupling parameter. It has to be noted that the origin acts as a {\emph{stable node}} within a smaller range of $\epsilon$ and also that one of the eigenvalues is nearly equal to zero. Hence, the origin behaves as a {\emph{stable star}} as shown in Fig. \ref{fig:13}(b). The bifurcation of the eigenvalues as a function of the coupling parameter is as shown in Fig.~\ref{fig:14}(a). In the $D^{*}_{\pm1}$ region, the origin is a {\emph{stable spiral}} for all values of the coupling strength. The {\emph{stable spiral}} nature of the origin in all the three regions of the difference system leads to asymptotic convergence of the trajectories around the origin in the synchronization sub-space.\\

Now we discuss the stability of the synchronized states observed in the coupled chaotic system using the {\emph{Master stability Function}} analysis. The phenomenon of complete synchronization (CS) is studied for two types of chaotic attractors observed at the control parameter values $f=0.1$ and $f=0.14$, respectively. Fig. \ref{fig:14}b(i) showing the MSF obtained for $x \leftrightarrow x^{'}$ coupling indicates that the coupled one-band (red line) and double-band chaotic attractors (blue line) entering into stable synchronized states and remaining thereafter for the coupling strengths $\epsilon \ge 0.02$ and $\epsilon \ge 0.04$, respectively. For coupled one-band chaotic attractors the Lyapunov exponents indicating hyperchaoticity are given as $\lambda_1 = 0.0709, \lambda_2 = 0.0395, \lambda_3 = 0, \lambda_4 = -0.4063, \lambda_5 = -0.4476, \lambda_6 = 0$ for $\epsilon = 0.005$. For coupled double-band chaotic attractors the Lyapunov exponents obtained for the coupling strength $\epsilon=0.015$ are given as  $\lambda_1 = 0.1599, \lambda_2 = 0.0517, \lambda_3 = 0, \lambda_4 = -0.3798, \lambda_5 = -0.4561, \lambda_6 = 0$. The existence of the x-coupled system in the stable synchronized states can be attributed to the {\emph{stable spiral}} nature of the origin in both regions $D^{*}_0$ and $D_{\pm1}$, respectively. A 3D plot indicating the variation of MSF as functions of the coupling parameter $\epsilon$ and control parameter $f$ is shown in Fig. \ref{fig:14}b(ii).\\
The numerical simulation results representing the unsynchronized and synchronized states of mutually-coupled {\emph{Murali-Lakshmanan-Chua}} circuits for x-coupled case are shown in Fig. \ref{fig:16}. The experimental observations confirming the analytical and numerical results given in Fig. \ref{fig:6} and \ref{fig:16} is shown in Fig. \ref{fig:17}.\\
\subsection{Forced Series LCR circuit with {\emph{simplified nonlinear element}}}
\label{subsubsec:2}

We present in this section, the synchronization dynamics of the mutually-coupled forced series LCR circuits having the {\emph{simplified nonlinear element}} as their nonlinear element. The sinusoidally forced series LCR circuit with a {\emph{simplified nonlinear element}} was introduced by Arulgnanam {\emph{et al}} \cite{Arulgnanam2009}. The normalized circuit parameters of this circuit take the values $\beta=0.9865,~\omega=0.7084,~a=-1.148,$ and $b=5.125$. The circuit exhibits a prominent chaotic attractor at the control parameter value $f=0.31$. \\

A linear stability analysis may be carried out for the difference system in each of the x-coupled piecewise linear regions based on the Jacobian matrix given in Eq. \ref{eqn:55}. In the $D^{*}_0$ region, the fixed point $(0,0)$ acting as a {\emph{saddle}} for $\epsilon < 0.074$ becomes an {\emph{unstable star}} in the range $0.074001 \le \epsilon \le 0.074002$,   and transforms into an {\emph{unstable focus}} in the range $0.074023 \le \epsilon \le 0.0807$ and becomes a a {\emph{stable focus}} for $\epsilon \ge 0.0808$ as the coupling parameter is increased. The bifurcation of the eigenvalues as a function of the coupling parameter is as shown Fig. \ref{fig:17}(a). In the $D^{*}_{\pm1}$ region the {\emph{origin}} is a {\emph{stable node}} for all values of $\epsilon$. The MSF corresponding to coupling of identical chaotic attractors obtained at $f=0.31$ indicates a transition of the coupled system into a stable synchronized state for $\epsilon > 0.04$ as shown in Fig. \ref{fig:17}(b). The numerical simulation results representing the unsynchronized and synchronized states of mutually-coupled systems are shown in Fig. \ref{fig:18}. The electronic circuit experimental results confirming the analytical and numerical results is shown in Fig. \ref{fig:19}.\\

\section{Mutually-coupled parallel LCR circuits: Numerical and Experimental results}
\label{sec:5}
In this section, we present the synchronization dynamics observed in mutually-coupled parallel LCR circuits. Two types of circuit systems differing by their constituent nonlinear element is discussed for $x \leftrightarrow x^{'}$ coupling of the state variables.
\subsection{{\emph{Variant of Murali-Lakshmanan-Chua}} Circuits}
\label{subsec:1}
The {\emph{Variant of  Murali-Lakshmanan-Chua}} circuit is a forced parallel LCR circuit with a {\emph{Chua's diode}} as the nonlinear element was introduced by Thamilmaran {\emph{et al}} \cite{Thamilmaran2000,Thamilmaran2001}. The normalized circuit parameters take the values  $\beta=0.05,~\omega=0.105,~a=-1.121,$ and $b=-0.6047$. This circuit exhibits significant chaotic attractors at two values of the control parameter $f$. The chaotic attractor at $f=0.39$ is obtained through the destruction of an invariant torus while that at $f=0.411$ is obtained through a reverse period-doubling sequence. The mutually-coupled system is studied for the complete synchronization of these two types of identical chaotic attractors. \\
Under $x \leftrightarrow x^{'}$ coupling, the Jacobian matrix of the difference system corresponding to the $D^{*}_{0}$ region is
\begin{equation}
J{_{0, \pm1}^{*}} =
\begin{pmatrix}
-1-2 \epsilon -
\begin{cases}
a, \text{$|x| < 1$}\\
b, \text{$|x| > 1$}\\
\end{cases}
&& -1 && \\
\beta && 0\\
\end{pmatrix},
\label{eqn:56}
\end{equation} 

A linear stability analysis of the difference system corresponding to the $D_{0}^{*}$ region shows that the origin is an {\emph{unstable spiral}} for $\epsilon < 0.0605$, a {\emph{stable spiral}} for $0.0605 \le \epsilon \le 0.284$ and becomes a {\emph{stable node}} for $\epsilon \ge 0.285$. In the $D_{\pm1}^{*}$ region, the origin transforms into a {\emph{stable node}} from a {\emph{stable spiral}} for $\epsilon \ge 0.026$. Figure \ref{fig:20}(a) and \ref{fig:20}(b) shows the trajectories of the difference system in the $x-y$ phase-plane for different values of the coupling parameter. The phase plot in Fig. \ref{fig:20}(a) shows the transformation of the {\emph{unstable spiral}} into a {\emph{stable spiral}} through the evolution of a {\emph{centre}}. Figure \ref{fig:20}(b) shows the trajectories of the {\emph{stable node}} nature of the origin in the $x-y$ phase plane. The bifurcation of the eigenvalues as function of the coupling parameter in the $D_{0}^{*}$ and $D_{\pm1}^{*}$ regions are shown in Fig. \ref{fig:21}(a) and  \ref{fig:21}(b), respectively. The MSFs for the $x \leftrightarrow x^{'}$ coupled systems are shown in Fig. \ref{fig:21}(b). Under $x \leftrightarrow x^{'}$ coupling, both the coupled identical chaotic attractors belonging to the control parameter values $f=0.39$ (red line) and $f=0.411$ (blue line) enters into stable synchronized states at the coupling strengths $\epsilon=0.0045$ and $\epsilon=0.004$, respectively and remain thereafter in the synchronized state as shown in Fig. \ref{fig:21}b(i). The variation of MSF as functions of the coupling parameter $\epsilon$ and $f$ are shown in Fig. \ref{fig:21}b(ii). \\
The numerical simulation results of the x-coupled systems indicating the unsynchronized states are  as shown in Fig. \ref{fig:22}. The corresponding experimental results are shown in Fig. \ref{fig:23}.

\subsection{Forced Parallel LCR circuit with {\emph{simplified nonlinear element}}}
\label{subsec:2}
The forced parallel LCR circuit with a {\emph{simplified nonlinear element}} exhibiting chaotic dynamics was introduced by Arulgnanam {\emph{et al.}}.\cite{Arulgnanam2009,Arulgnanam2015} The synchronization dynamics of the unidirectionally-coupled circuit has been studied analytically, numerically and experimentally.\cite{Sivaganesh2017} The normalized state equations of the circuit are as given in Eq. \ref{eqn:6} with the system parameters taking the values $a=-1.148,~b=5.125,~\beta=0.2592$ and $\omega=0.2402$. The system exhibits significant chaotic attractors at two values of the control parameter $f=0.695$ and $f=0.855$, respectively. The Jacobian matrix of the difference system corresponding to $x \leftrightarrow x^{'}$ coupling is as given in Eq. \ref{eqn:56}. A linear stability analysis in the $D_{0}^{*}$ region reveals the transformation of the origin from an {\emph{unstable spiral}} into a {\emph{stable spiral}} with the evolution of a {\emph{centre/elliptic}} in the range $0 \le \epsilon \le 0.583$ and becoming a {\emph{stable node}} for $\epsilon > 0.583$. The bifurcation of the eigenvalues as a function of the coupling parameter is as shown in Fig. \ref{fig:24}(a).  However, in the $D_{\pm1}^{*}$ region, the origin acts as a {\emph{stable node}} for all values of the coupling strength. The MSF in the $x \leftrightarrow x^{'}$ coupled case shows the chaotic attractors pertaining to $f=0.695$ and $f=0.855$ enters into stable synchronized states at the coupling strengths $\epsilon=0.04$ and $\epsilon=0.06$, respectively, as indicated by the negative values of MSF shown in Fig. \ref{fig:24}b(i). Beyond this lower bound of synchronization the couple systems exist in the synchronized state forever, as the coupling strength is increased. The variation of MSF as functions of $f$ and $\epsilon$ is shown in Fig. \ref{fig:24}b(ii). \\
The phenomenon of complete synchronization observed in the x-coupled systems through numerical simulation and experimental observations are shown in Fig. \ref{fig:25} and \ref{fig:26}, respectively.

\section{Summary and Conclusions}
\label{sec:6}

We have developed in this paper, explicit generalized analytical solutions to a class of second-order, non-autonomous simple electronic circuit systems with two different nonlinear elements. Explicit analytical solutions for the forced series and parallel LCR circuit systems with different nonlinear elements were presented and their synchronization dynamics were studied using the solutions. The analytical results thus obtained have been validated through numerical simulation results. The stability of the synchronized states were analyzed using the MSF approach. By analyzing the eigenvalues of the difference system in each of the piecewise linear region, we could arrive at a sufficient condition for complete synchronization in the mutually-coupled simple second-order chaotic systems. The existence of stable fixed points in any one of the piecewise linear regions of the difference system guarantee synchronization of the coupled systems. Further, the numerical and analytical results are confirmed through electronic circuit experimental results for each of the circuit systems. The phenomenon of complete synchronization in mutually-coupled systems have been reported through analytical results in the literature of the first time. Following the solutions to the normalized state equations of quasi periodically forced and unidirectionally coupled simple chaotic systems, the analytical solutions to mutually-coupled systems is developed which could be further developed for analyzing simple chaotic networks involved in secure transmission of signals.

\section{Acknowledgement}

One of the authors A. Arulgnanam gratefully acknowledges Dr.K. Thamilmaran, Centre for Nonlinear Dynamics, Bharathidasan University, Tiruchirapalli, for his help and permission to carry out the experimental work during his doctoral programme.

\end{document}